\documentstyle[aps,floats]{revtex}
\begin{document}
\input epsf
\renewcommand{\topfraction}{1.0}
\twocolumn[\hsize\textwidth\columnwidth\hsize\csname
@twocolumnfalse\endcsname

\title{High energy protons from PKS 1333-33}
\author{L. A. Anchordoqui $^a$, G. E. Romero $^{a,b}$, S. E.
Perez Bergliaffa $^c$ and J. A. Combi $^b$}
\address{$^a$ Departamento de F\'{\i}sica, UNLP, C.C. 67, (1900) La Plata
Argentina}
\address{$^b$ Instituto Argentino de Radioastronom\'{\i}a, C.C. 5, 
(1894) Villa Elisa,
Argentina}
\address{$^c$ LAFEX/CBPF-Centro Brasileiro de Pesquisas Fisicas, Rua Xavier Sigaud 150,
Rio de Janeiro, 22290-RJ, Brazil}
\maketitle

\begin{abstract}

In this letter we give an account of the possible acceleration of protons
in the outer radio lobes of the active galaxy PKS 1333-33. We also 
make estimates of the arrival energy spectrum. 

\noindent {\it PACS number(s):} 96.40 - 98.70.S - 95.85.R  - 13.85.T
\end{abstract}
\vskip2pc]

Radio galaxies considered as sources of cosmic rays (CR) beyond the 
Greisen-Zatsepin-Kuz'min cutoff \cite{gzk}
must be quite close to our own galaxy ($z \leq 0.03$). There are just a few
objects within this range and, consequently, the expected ultra high energy
CR distribution on
the sky should be highly anisotropic. Assuming that the intergalactic
magnetic field is near 1 nG, it should be expected an excess of CR
detections at energies larger than $5\times10^{19}$ eV within regions of
angular radius $\theta\leq 15^{\circ}$ and centered at the positions of
the nearest active galaxies \cite{jeremy}.
According to this picture, the southern
CR-sky should be dominated by three outstanding sources: Cen A (the
nearest radio galaxy) which would provide the most energetic particles
detectable on Earth \cite{cena}, Pictor A (a strong source with a flat
radio spectrum) which would contribute with the larger CR flux
\cite{rabir}, and PKS 1333-33 which might be a source of events similar to
those recently detected in the Northern Hemisphere. In addition to these
sources, there are other two southern candidates, Fornax A ($z=0.057$) and PKS
2152-69 ($z=0.027$), which
could provide contributions to the CR flux above the cutoff. An approximate
theoretical picture
of the up to now unexplored ultra high energy CR southern sky is
consequently at our disposal.
The present letter is devoted to CR production in the outer region of the
nearby southern active galaxy PKS 1333-33
and the spectral modifications arisen from their propagation through 
intergalactic space.

The radio galaxy PKS 1333-33 is made up of a core, two 
symmetric jets, and two extended radio lobes. Fig. 1 shows the entire source
at 20 cm according to the VLA map obtained by Killeen et al. \cite{kil1}. 
The core
has been identified with the E1 galaxy IC 4296, which has a redshift $z$=0.013 
\cite{slee}. The distance to the source is 35.2 $h^{-1}$ Mpc, 
($h=H_0/100$ km$^{-1}$ s Mpc, hereafter we shall adopt $h
= 0.65$ \cite{freedman}). The large-scale structure of PKS 1333-33 has been
studied in detail at 1.3, 2, 6, and 20 cm with an 
angular resolution of 3.2'' \cite{kil1}. The jets are slightly bent, 
presumably as a consequence
of the motion of the core with respect to the intergalactic medium.
The total flux density of the source at 20 cm is $\sim$ 14.5 Jy. The integrated
radio luminosity is $\sim 2.5 \times 10^{41}$ erg s$^{-1}$
assuming a spectral index of $\alpha = 0.6$ and frequency cutoffs at
$10^7$ and $10^{11}$ Hz. Actually, the value $\alpha = 0.6$ is correct
only for the extended radio lobes. The spectral index steepens in the 
jets, reaching $\alpha =1$, while the core has flatter values: 
$\alpha \sim 0.3$.

The hot spots of extended radio sources are regions of strong synchrotron
emission \cite{mei}. These regions are produced when the bulk
kinetic energy of the jets ejected by a central active source (supermassive
black hole plus accretion disk) is reconverted into relativistic particles
and turbulent fields at a ``working surface'' in the head of the jets
\cite{blare}. The speed $v_{\rm h}$ with which the head of a jet advances
into the intergalactic medium of particle density $n_{\rm e}$ can be obtained 
by balancing the momentum flux in the jet against the momentum flux of the 
surrounding medium. Measured in the frame comoving with the advancing head,
$v_{\rm h}\approx \;v_{\rm j}\,[ 1 + ( n_{\rm e} /n_{\rm j})^{1/2}]^{-1}$,
where $n_{\rm j}$ and $v_{\rm j}$ are the particle density and the velocity
of
the jet
flow, respectively. Clearly, $v_{\rm j}> v_{\rm h}$ for 
$n_{\rm e} \geq n_{\rm j}$, in such a way that the jet will decelerate. The 
result is the formation of a strong 
collisionless shock, which is responsible for
particle reacceleration and magnetic field amplification. 
The acceleration of particles up to ultrarelativistic energies in the hot
spots is the result of repeated scattering back and forth across the shock
front. The particle deflection in this mechanism is produced by Alfv\'en waves 
in the turbulent
magnetic field. Biermann and Strittmatter \cite{bistri} have studied this 
process assuming that the energy density per unit of wave number of MHD 
turbulence is of Kolmogorov type. According to their calculations,
the highest energy of protons injected in
the intergalactic medium from the
hot spot can be obtained by balancing the gains and losses in the diffusive
shock acceleration process:
\begin{eqnarray}
E_{{\rm max}} & = & 7.8\times 10^5 \;\beta_{_{\rm jet}}\;
\!\!\!\!\!\!\!^{\;3/2}
\;u^{3/4}\;R_{_{\rm hs}}^{\;\;-1/2} \\ \nonumber  
 & \times & B_{-5}^{-5/4} \; ( 1 + A\,a)^{-3/4} \;\,{\rm EeV}
\label{emaxim}
\end{eqnarray}
where  $\beta_{_{\rm jet}}$ stands for the jet velocity in units of $c$, $u$
is the ratio of turbulent to ambient magnetic energy density in the hot spot
(of radius $R_{\rm hs}$ 
measured in kpc), $B_{-5}$ is the magnetic field in units of $10^{-5}$ G, $a$
is the ratio of photon to magnetic energy density, and $A$ gives a relative
strength of $\gamma p$ interactions against the synchrotron emission.

An interesting feature of PKS 1333-33 is an intense region of synchrotron
emission localized at the outer edge of the eastern lobe. This region can be
considered as a ``working surface'' formed by the deceleration of the jet.
This interpretation is supported by VLA polarimetric observations,
which show a change in the field orientation
from parallel to perpendicular to the jet axis  \cite{kil1}.
This change is probably due to the rearrangement of the field lines in
the post-shock region. The degree of linear polarization in the eastern lobe
is in the range 20\%-40\%. The synchrotron parameters
estimated for this region are: 
the minimum energy density ($\epsilon_{\rm min} = 2.1\times 10^{-13}$ 
erg cm$^{-3}$), the  
minimum magnetic field ($B_{\rm min} = 1.5\times 10^{-6}$ G),   
the minimum total pressure ($P_{\rm min} = 1.1\times 10^{-13}$
dyne cm$^{-2}$), and the 
degree of linear polarization ($m$ = 20-40\%).

The leptonic component of the CRs will
produce synchrotron emission with a spectrum given by 
$S_\nu\propto \nu^{-\alpha}$, where $\alpha = (\Gamma -1)/2$.
Since $\alpha = 0.6$, we get $\Gamma = 2.2$. In what follows, we
shall assume that electrons and protons in the source obey the same
power law energy spectrum $\propto E^{-\Gamma}$.

\begin{figure}
\label{fpks}
\centering
\leavevmode\epsfysize=7cm \epsfbox{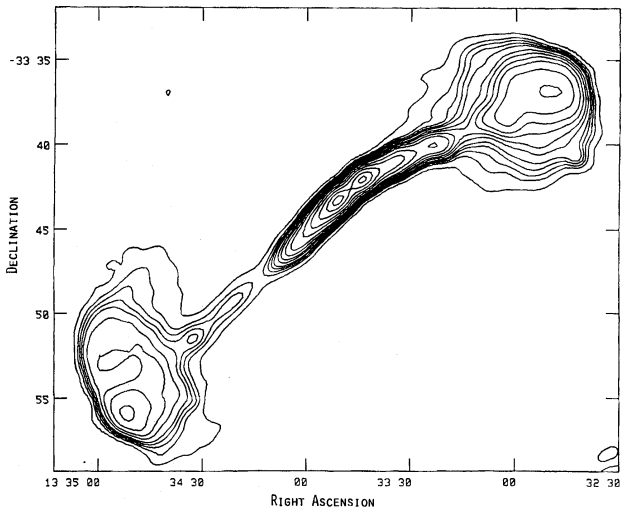}\\
\caption{VLA image of the entire PKS 1333-33 source obtained at a 
wavelength of 20 cm by Killeen et al.} 
\end{figure}

The degree of linear polarization expected for the synchrotron radiation
when the magnetic field is homogeneous is
\begin{equation}
m_0(\Gamma)={3\Gamma +3\over{3\Gamma +7}}\approx 71\%
\label{homo}
\end{equation}
However, the observed degree of polarization has a mean value of $m(\Gamma)\sim
30\%$. This fact can be explained by the presence of a turbulent
component $B_{\rm r}$ in the field, in such a way that
\begin{equation}
m(\Gamma)=m_0(\Gamma)\;{B_0^2\over{B_0^2+B_{\rm r}^2}}
\label{ran}
\end{equation}
where $B_0$ stands for the homogeneous field. From Eq. (\ref{homo}) and
(\ref{ran}) we get
\begin{equation}
B_{\rm r}\approx 1.2\;B_0
\end{equation}
and consequently $u=B^2_{\rm r}/B^2_{\rm total}\approx 0.6$.

The radius of the acceleration region can be directly measured by means of 
a Gaussian fitting from the
detailed VLA maps obtained by Killeen et al. \cite{kil1}, resulting $R_{\rm hs}
\approx 2.5\;h^{-1}$ kpc. The velocity of the jet is not well established. 
If the source is
$\sim 10^7$ yr old, a velocity $\beta_{\rm jet}\sim 0.2$ can be estimated from
an analysis of the energy budget \cite{kil2}. The value of $A$
in Eq. (\ref{emaxim}) has been computed by Biermann and Strittmatter \cite{bistri}
almost independently of the source parameters. They obtain $A\approx 200$. A
value $a\sim0.01$ seems to be reasonable for a source with the 
luminosity of PKS 1333-33 \cite{bistri}. Taking the above considerations 
into account (with a typical value for the total 
magnetic field in the spot of $B_{-5}\sim 10$ \cite{biermann}), we obtain 
from Eq. (\ref{emaxim}) the maximun injection energy for protons, 
$E_{{\rm max}} \approx 6 \times 10^{20}$ eV. 

The diffusive shock acceleration 
process in the working surface of the eastern
lobe leads to a power law particle spectrum \cite{bell},
\begin{equation}
Q(E)= \kappa \; E^{-\Gamma}\;\;  (E_0< E <E_{{\rm max}})
\label{espec}
\end{equation}
where $\kappa =(n_0/\Gamma -1)\;E_{0}^{1-\Gamma}$ for $E_{\rm max}\gg
E_{0}$,  
and $n_0$ is the particle density in the source. Thus, we have the
following proton injection spectrum
\begin{equation}
Q(E) \propto E^{-2.2},\;\;\;\,E_{0}< E <6 \times 10^{20}\;{\rm eV}.
\label{pacota}
\end{equation}

Let us now estimate the losses that the high energy protons
suffer during their journey through the intergalactic medium, 
and the consequent modification of the injection spectrum.
This energy degradation originates in
the interaction of the protons with the very low energy photons of the 
cosmic microwave background (CMB). For nearby
sources (distances $<$ 100 Mpc)
photomeson production is the 
dominant interaction  
at $E > 10^{19}$ eV (if one considers the contribution coming from 
the tail of the Planck distribution) \cite{prd1}.
The propagation of the protons through the relic gas of photons may be
modelled by a kinetic transport equation.
Although it is clear that the interactions of the ultra high energy protons 
with the CMB photons lead to a step-by-step energy loss (which should be
included in the transport equation as a collision integral), 
hereafter we shall use the continuous energy loss approximation
assuming straight line propagation which is expected to be
in agreement with the physics of the problem at hand. 

Using the formalism presented in \cite{prd1} it is straightforward to
compute the main
characteristic of the evolved spectrum.
It is well known that the conservation of the differential particle 
number implies that,
\begin{equation}
\frac{\partial N}{\partial t} = \frac{\partial (b(E) N)}
{\partial E} + D \,\, \nabla^2 N + Q,
\label{@}
\end{equation}
where, in the first term  on the right, $b(E)$ is the mean rate at which particles 
lose
energy ($\approx 3.66 \times 10^{-8} \, E \, 
\exp\{287/E\}$ EeV/yr \cite{prd1}); 
the second term, the diffusion in the CMB,
is found to be extremely small due to the low density of relic
photons and is neglected in the following.
The third term corresponds to the particle injection rate into
the intergalactic medium given by Eq. (\ref{pacota}).
The solution of Eq. (\ref{@}) reads,
\begin{equation}
N(E,t) = \kappa \; E_{g}^{-\Gamma}\; {b(E_{g})\over{b(E)}}
\end{equation}
with the injection energy $E_{g}$ fixed by the constraint,
\begin{equation}
3.66 \times 10^{-8} \,t \,{\rm yr}^{-1} - {\rm Ei}(287 
/E) + {\rm Ei}(287/E_g) = 0
\end{equation}
being Ei$(x)$ the exponential-integral function.
The evolution of the injection
spectrum can be then conveniently represented by the modification 
factor $\eta$, defined as the ratio between the modified and the unmodified
spectrum. In Fig. 2 we have plotted the relation obtained for $\eta$ in
the case of
PKS 1333-33. The energy loss by photomeson production creates the
expected cutoff, and the resulting ultra high energy CRs (protons and
neutrons) pile up just
below the threshold energy of photopion production, forming a bump.
However, the energies at which the cutoff and the bump appear seem to
indicate that CRs above 100 EeV might be expected from the location in the 
sky of PKS 1333-33 ($l \approx 313.7^\circ$, $b \approx 27.7^\circ$).

The question whether this scenario is the correct one or not should be
answered in a few years by the Pierre Auger Southern Observatory \cite{auger}
(fluorescence detector plus ground array) as well as by the future eyes of
the OWL \cite{owl} that will deeply watch into the CR-sky.

\begin{figure}
\label{}
\centering
\leavevmode\epsfysize=7cm \epsfbox{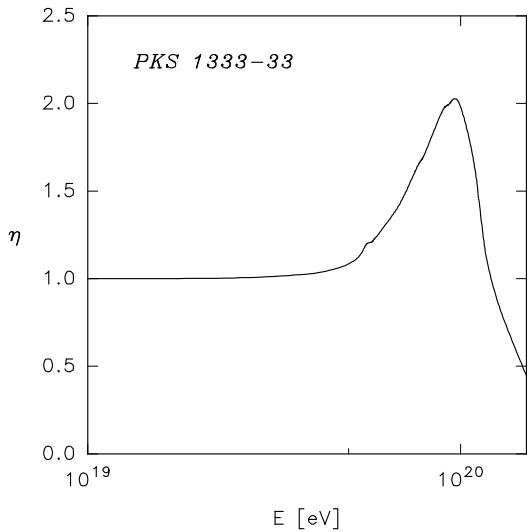}\\
\caption{Modification factor of PKS 1333-33.}
\end{figure}

\acknowledgements

We are indebted to N. E. B. Killeen,
G. V. Bicknell, and R. D. Ekers for permission to
reproduce the VLA image of PKS 1333-33. G. E. R. especially thanks the
help by J. Blumina Romero. This work has been partially
supported by FOMEC, CONICET, CLAF-CNPq, and the agency ANPCT.

\end{document}